\documentclass[prepint]{elsarticle}
\usepackage{graphicx}

\begin{document}

\title{Development of an IMS Type Device for Volatile Organic Compunds
  Detection: Simulation and Comparision of the Ion Distributions
}
\author[FaCENA]{Guillermo P. Ortiz\corref{cor1}}\ead{gortiz@exa.unne.edu.ar} 
\cortext[cor1]{Corresponding author}
\author[CAC]{Carlos A.Rinaldi\fnref{fn1}}
\author[CAC]{Norberto G. Boggio\fnref{fn1}}
\author[CAC]{Juan Vorobioff}
\author[CAC]{Juan J.  Ortiz}
\author[FaCENA]{S. G\'omez\fnref{fn1}}
\author[FaCENA]{G.A. Aucar\fnref{fn1}}
\author[CAC]{A. Lamagna}
\author[CAC]{A. Boselli}
\fntext[fn1]{Fellow of CONICET}
\address[FaCENA]{Departamento de F\'isica, Facultad de Ciencias Exactas, Universidad Nacional del Nordeste,
  Av. Libertad 5400 Campus-UNNE, W3404AAS Corrientes, Argentina. %
}
\address[CAC]{Instituto de Nanociencia y Nanotecnolog\'ia, Comisi\'on Nacional
  de Energ\'ia At\'omica, CAC Av. Gral Paz 1499 San Mart\'in , Buenos Aires,
  Argentina.%
}


\begin{abstract} 
  Ion Mobility Spectrometry (IMS) is a well-known, sensitive and rapid
  technique to detect dangerous organic compounds. We propose a system in
  which a crown type discharge generates a ionic flux that is swept towards an
  array of collectors by a transverse electric field.  The ions are separated
  as they enter the cell according to their mobility.  Thus, the distribution
  of the charge collected at the detector assembly constitutes a {\em
    fingerprint} for each organic compound. Simulations of our cell and
  experiments were performed for small amount of Acetone, Ethanol and
  Toluene. The dependence on the cell parameters of the current and charge
  versus time of flight was analyzed. Our simulation reproduces only
  qualitatively the experimental results. However, a PCA statistical analysis
  of the simulated results shows that such a fingerprint can be used to
  identify those compounds with certainty.
\end{abstract}

\maketitle

\section{Introduction}

The technique for identification of volatile organic compounds by differences
of ion mobility is one of the most efficient techniques applied for detection
of explosives, drugs and industrial toxic compounds~\cite{Moore(2004)}. The
detection limit of this technique is found to be of the order of $10^{-9}$
(ppb). It has a very rapid response time, which makes it very attractive for
situations of high control demands.  The ions of the molecule to be identified
acquire a velocity under the influence of an external electric field $\vec
E$. At the lowest order

\begin{equation}
  \label{eq:K_def}
  \vec v_E=K \vec E,
\end{equation}
where $K$ is the mobility. If the field is intense enough it has been
reported~\cite{Revercomb(1975)} that one should go two orders higher (cubic) in
the field dependence. A way to understand this result is by considering a
power series expansion on $\vec E$. Assuming isotropy of $K$ and applying
symmetries on spatial transformations, the vectorial properties of $\vec v_E$
and $\vec E$ determines that only odd terms in such expansion are possibles.

Besides the interaction of the ions with the field $\vec E$, they collides
with molecules of the carrier gas that flow through the detection system under
a laminar flux. The explicit dependence of $K$ on the pressure due to the
carrier gas, the temperature, the ion charge, its mass and the cross section
can be obtained from general considerations when the field $\vec E$ is not
very intense~\cite{Revercomb(1975)}. Working on the center of mass framework
to describe the collisions between two bodies, and the reduced mass $m_r=M
m/(M+m)$ where $m$ is the mass of ions and $M$ the mass of the molecules of
the gas, the variation of the linear momentum of ion-gas system is due to the
external impulse of force, {\it i.e.} $q E \tau$, being $q$ the ion charge and
$\tau$ the characteristic time between collisions. It can thus be established
that

\begin{equation}
  \label{eq:v_E}
  v_E=q E\tau/m_r.
\end{equation}
Furthermore $\tau$ can be related to a collision cross section of the ion-gas
system $\Omega$ and the density of this system, which is approximately equal
to the gas density $\eta$. Considering the characteristic length $l=\tilde
v\tau$ and the average rate $\tilde v$ of the reduced system ion-gas between
collisions, it is estimated that in the volume $l \Omega$ a collision should
typically occur. Hence $\eta\approx1/(l \Omega)$ and Eq. (\ref{eq:v_E}) can be
written

\begin{equation}
  \label{eq:v_Ebis}
  v_E=\frac{q E}{m_r}\frac{1}{\eta\Omega\tilde v}.
\end{equation}
One can approximate $\tilde v\approx(3k_BT/m_r)^{1/2}$ because the kinetic energy of the ion-gas system is mainly due to thermal agitation of value 
$k_BT/2$ for each degree of freedom. Then 

\begin{equation}
  \label{eq:v_Efinal}
  v_E=\frac{q E}{\eta\Omega}\left(\frac{m+M}{mM3k_BT}\right)^{1/2},
\end{equation}
being $k_B$ the Boltzman constant and $T$ the absolute temperature. In
Eq. (\ref{eq:v_Efinal}) there is a factor $3^{-1/2}$ that should be replaced
by $3(2\pi)^{1/2}/16$ when more advanced molecular calculations are
considered~\cite{Jurs(1996)}.

We include such derivation here because it contains the right dependence with
the physical parameters of interest and its derivation is straightforward. In
the calculations presented in this work we employ instead the mobility usually
reported in the literature ~\cite{Agbonkonkon(2004),Appelhans(2005)}, which is
obtained from

\begin{equation}
  \label{eq:K}
  v_E/E=\frac{3q}{16\eta\Omega}\left(\frac{2\pi(m+M)}{m M k_BT}\right)^{1/2}.
\end{equation}
This mobility is standardized at normal $P_0$ pressures and $T_0$ temperatures
through
\begin{equation}
  \label{eq:K_0}
  K_0=K(T_0/T,P/P_0),
\end{equation}
$T_0=273~^{o}\!K$ and $P_0$ depends on the system of unities used to refer to
1013 HPa. $K_0$ is known as the reduced mobility.

A decade ago the identification of species by separation
techniques~\cite{Eiceman(1999)} was based on the differences in the reduced
mobilities given by Eq.(\ref{eq:K_0}). Since then substantial modifications
have been made in the design and size of sensor devices. Using asymmetric
electric fields in the range of radio-frequency and by micro-machining the
detection cell, a field asymmetric or FA-IMS variant was
proposed~\cite{Miller(2000)}. Identifications of up to 0.1 ppm of Toluene and
also a resolution threshold of 60 ppb for a mix of gases of Acetone-Benzene
and Acetone-Toluene were reported. It has been recently
proposed~\cite{Gillig(2004)} a combination of mass spectroscopy and ion
mobility to improve the transmittance in an Ion Mass Mobility Spectroscopy
through a focus cell by means of periodical electrostatic fields distributed
on a very long arrangement of electrode rings (1m of length). It has also been
proposed recently an IMS type cell by swept-field scanning~\cite{Solis(2006)}
in which a plane and parallel detectors system produces an electric field
perpendicular to the flux of the ions being detected. The operating principle
of this device for mobility identification is based on the differences of the
average path described by the ions within the detection cell.
\begin{figure}[h]
  \centering
  \label{fig:barrido}
\includegraphics{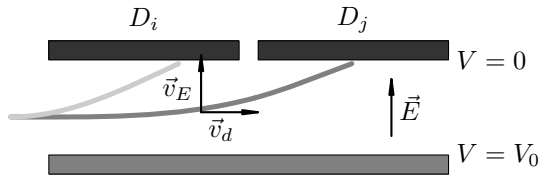}
\caption{Average paths within the detection cell. $D_i$, $D_j$ represent two
  of the seven detectors contained in the cell. The detectors are earthed and
  the anode is kept at $V_0$ potential. The left path corresponds to an ion
  with higher mobility with respect to the right path.}
\end{figure}

In Fig.~\ref{fig:barrido} an example of two ion paths corresponding to ions
with different mobility is given. Due to the velocity of the carrier flux, the
ions reach a detection zone with velocity $\vec v_d$ parallel to the detectors
plane. By the action of the transversal field $\vec E$, the ions acquire,
according to Eq. (\ref{eq:K_def}), a vertical component velocity $\vec v_E$
proportional to the mobility $K$. By adjusting the geometrical parameters of
the cell, the flux of the carrier gas and the potentials among electrodes, it
is possible to obtain a typical ion distribution for each compound on the
$D_i$ and $D_j$ detectors, where $i\ne j=1\ldots n$ are labels. This is
characterized by a greater accumulation of charges on the detectors, whose
positions measured from the gas input in the cell are more distant for ions
with less mobility, as is suggested schematically in Fig.~\ref{fig:barrido}.

A modification in the swept-field scanning detection was recently proposed.
It includes an ion-focusing just before the introduction of ions in the
detection cell~\cite{Zimmermann(2007)}. In this variant, the flux of both the
carrier gas and the ionized gas are separated through geometrical
constrictions. This increase the precision of the positions reached by the
ions in the detection cell. In that sense, the aim of such a proposal is not
meant to obtain a fingerprint for each compound. Indeed it proposes that the
identification shall be determined by the average time taken by each species
to reach a specific detector~\cite{Zimmermann(2007)}. Although the ion time of
flight has been analyzed for different values of carrier gas flux and
potentials between electrodes in order to evaluate the focalization
conditions, it has been found in this work that identification of species by
statistical analysis of their traces on detectors is more accurate if few
restrictions are imposed by operational conditions, {\it i.e.}, work pressure
and non-radioactive ionization method.  Furthermore, due to the fact of a very
high difference of electric potential generated by the crown effect, fragments
of the molecule to identify are produced, each one with a different
mobility. In our work we have found that this effect is consistent with the
formation of a fingerprint for each species at the moment of measuring the
total charge deposited on the detector assembly.

Basically, the main idea behind a fingerprint identification is the following:
A system is designed in such a way that traces of given chemical components
are well defined under specific operating conditions. With this purpose,
numerical simulations are first performed, which allow us to identify these
specific conditions. The algorithm used to simulate the ion detection system
has proven to be sufficiently robust for the design of ion mobility detection
cells~\cite{Lai(2008)}. Furthermore in our work, the design of an experimental
prototype is analyzed for obtaining traces of Ethanol, Acetone and
Toluene. The statistical study of these traces was performed by the principal
components analysis~\cite{Pearson(1901)}, known as PCA (acronyms from
Principal Components Analysis), which was used earlier on similar
identifications within the design of an electronic
sniffer~\cite{Lamagna(2004)}.

We present our work in the following way. In Section~\ref{sec:exp} we
describe the experimental configuration of the prototype device and the
procedure we followed for the measurement of signals in the identification of
three different chemical species. In Section~\ref{sec:simu} we present a brief
description of the algorithm used for the simulation of ion paths. We divided
Section~\ref{sec:results} in four parts in order to show the experimental
results in Subsection~\ref{subsec:exp}, time of flight simulations in
Subsection ~\ref{subsec:tof} and collected charges in Subsection
~\ref{subsec:ca}. In Subsection ~\ref{subsec:pca} we present the analysis of
principal components and Section ~\ref{sec:conclusion} is devoted to the main
conclusions.

\section{Experimental device}\label{sec:exp}

\begin{figure}[h]
  \centering
  \includegraphics[scale=0.35]{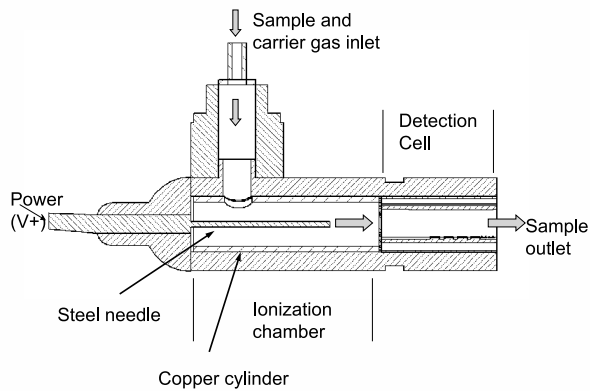}
  \caption{Lateral section of the experimental device. The upper part 
    corresponds to the carrier gas inlet. In the central part ionization 
    is produced, and the detection cell is on the right. %
  }
  \label{fig:esquemaExp}
\end{figure}

An experimental device has been designed as indicated in
Fig.~\ref{fig:esquemaExp}. It consists of three principal parts: i) the
carrier gas flux inlet, ii) the ionization chamber and iii) the detection
cell. The sample is
introduced as a gas in the ionization chamber together with the carrier gas
where, through a type of crown discharge, is ionized. The pointer shaped
electrode, shown in the central part of Fig.~\ref{fig:esquemaExp}, is a
stainless steel needle of 0,8 mm of diameter fixed inside a teflon
cylinder. An earthed copper tube of 8 mm of diameter closes the discharge
circuit of the pulse. The ionization discharge is recorded with an
oscilloscope (Tektronix P6015). The needle is concentric to the copper tube;
hence the distance between electrodes is approximately 4 mm. A stainless steel
mesh is placed between the ionization chamber and the detection cell in order
to screen the high field produced during the ionization discharge within that
last zone. The detection arrangement is similar to that proposed by Solis {\it
  et al.} \cite{Solis(2006)}; the field $\vec E$ is transversal to the ion
flux determined by the carrier gas. Our ion detection device is an assembly of
7 earthed detectors placed in a plane as shown in Fig.~\ref{fig:esquemaExp}
and connected to a constant potential which is fixed in the range between 10
and 100V. The dimensions of the detection cell are 17mm $\times$ 11mm $\times$
2mm. A dry carrier gas of N$_2$ is used in order to avoid humidity problems in
the air (Indura 99,998\%, H$_2$O $\le$ 3mg/ml). The gas flux is controlled by
a flux measuring device: Brooks model 6-658. The testing organic substances
are absorbed in a piece of cotton placed in a glass tramp. A flux of N$_2$ gas
passes through the tramp dragging the steam of the sample at room temperature
and directed to the ionization zone. The ions generated are dragged by the
same flux of N$_2$ gas towards the detection cell. Each electrode is connected
to a resistance of 3.2 M$\Omega$ and the electrical signal generated by the
ions is measured in a oscilloscope Instek GDS 810S synchronized with the pulse
of the discharge of the ionization needle. All the system is controlled with a
PC and due to the high signal/noise relationship, only 8 signals are necessary
to average the response of each detector. With this system it is possible to
study: Absolute Ethanol (Merck, pro-analysis, 99.8\%), Toluene (Mallinckrodt,
analytical, 99.8\%) and Acetone (Sintorgan, analytical, 99.5\%). All of them
were used without any previous treatment.

\section{Numerical simulations}\label{sec:simu}

The ion paths in the detection cell were simulated with the SIMION 8.0 code.
The SDS module was implemented. This module combines the dynamics of a
particle in a viscous medium and its interaction with the carrier gas at room
pressure~\cite{Appelhans(2005)}. We give here a brief description of such a
simulation. In order to define an ion path, the equation of motion is
considered, in which acceleration has a viscosity component due to velocity in
Eq.(\ref{eq:K_def}), and another Coulombic component due to the electrostatic
potential of the cell. Furthermore when the SDS module is implemented, a
diffusion model is incorporated. The statistics of the distance covered after
10$^5$ collisions in 10$^5$ ion samples was applied. From the ion-gas mass
relationship data of 1, 10, 100, 1000 and 10000, the distance corresponding to
intermediate values of mass were calculated by interpolation. Such distance is
used to modelling the size of the jump in a step of the path. The direction
within the 3D space of the cell should be calculated with a uniform
distribution in order to modelling the expected local isotropy. In this sense,
an appropriate sample space should be chosen. For instance, if a point within
a cube is randomly taken in order to obtain a random direction, the
appropriate sample space is defined only by those points that belong to the
sphere inscribed in such a cube~\cite{ErrataSDS}. Once the path is defined,
the algorithm reports the positions of the ion's impact. We have instrumented
a couple of codes written in interpretative language (Perl) to process the
data accompanying each ion path. With these codes we obtain the charge
collected as a function of time. Through a numerical derivation of the
collected charge, the current in each detector as a function of the time of
flight was obtained.

The SIMION 8.0/SDS algorithm~\cite{Appelhans(2005)} allows us to estimate
(from the mass data) or to take as a predeterminated information, the
mobilities of the ions whose paths will be modeled. Two types of ion path
simulations were considered: one without fragmentation and the other with
fragmentation. This means that, for each species studied, ions of the whole
molecules were considered first, and secondly, the fragments of such
molecules according to their relative abundances~\cite{Nist(2005)} being the
least of them higher than 10\%.
\begin{figure}[h]
\begin{center}
\includegraphics[width=12cm]{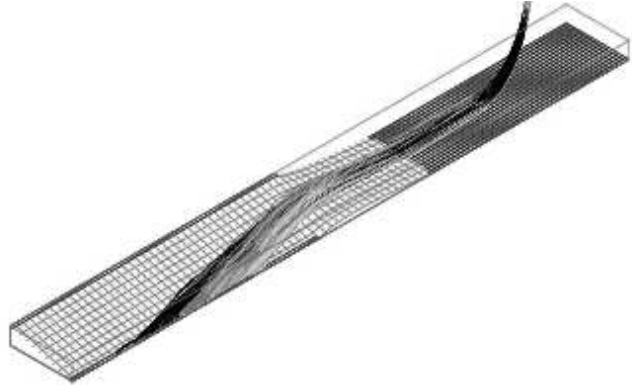}
\caption{\label{fig:juntos} Surface potential energy view of the ion path
  simulation of a Toluene (black) - Acetone (light gray) - Ethanol (gray)
  mixture obtained by Simion 8.0. The Ions come from the rigth top side of
  figure. The top part of the traces correspond to the potential of needle
  discharge (5000V).  The carrier flux with velocity 30 m/s allows the flow of
  the ions go into detector zone where potential increments are indicated by
  the surface potential level. In the left bottom side of figure ions
  impacts to detector at 0V at the lowest part of the surface potential. The
  anode is opposite to it in the upper part of the surface potential at
  50V.%
}
\end{center}
\end{figure}
Fig.~\ref{fig:juntos} shows the surface potential energy view obtained with
Simion 8.0 for the simulation of paths considering only primary ions (without
fragmentation) of Toluene, Acetone and Ethanol.  Results indicate that the
ions of Acetone and Ethanol with greater mobility are collected preferentially
in the detectors closer to the capture cell inlet (see also
Fig.~\ref{fig:esquemaExp}).

\section{Results}\label{sec:results}

In this section, experimental results and numerical simulations of the device
proposed here for the detection cell by swept-field scanning are
presented. Flight times are analyzed as well as the collected charges in
detectors for different combination of carrier gas flux and detector
potentials. The analysis of the principal components for one of the selected
combinations is also presented. The experimental results are given in the
first subsection, and in the following subsections the results of simulations
and numerical calculations are shown.

\subsection{Experimental}\label{subsec:exp}

\begin{figure}[h]
  \centering
  \includegraphics{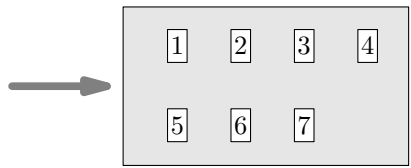}
  \caption{Diagram that shows the numbering for the array of detectores. Ions
    flux goes into detection cell from the left as indicated by the arrow.%
  }
  \label{fig:detectores}
\end{figure}

\begin{figure}[h]
  \centering
  \includegraphics{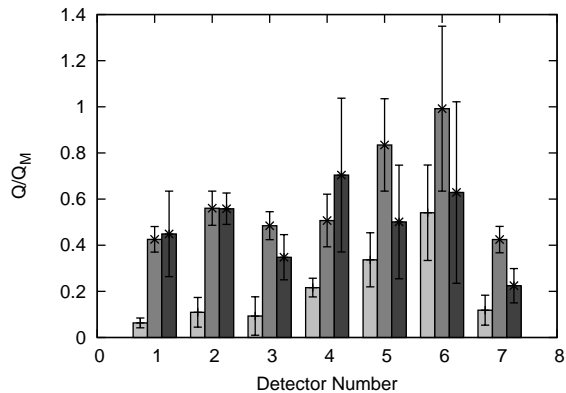}
  \caption{Collected charge measured in each detector when Acetone (light
    gray), Ethanol (gray) and Toluene (dark gray) samples are tested
    separately.%
  }
  \label{fig:DetAllExp}
\end{figure}

Under normal conditions of pressure and temperature different tests were
performed as described in Section~\ref{sec:exp} for Acetone, Ethanol and
Toluene at saturated steam pressure using the device shown in
Fig.~\ref{fig:esquemaExp}. The location of the detectors in the detection cell
is shown in Fig.~\ref{fig:detectores}.
Ion gas moves from left to right as indicated by the arrow. The charge
collected during the detection process was obtained for each detector.  In
Fig.~\ref{fig:DetAllExp} these data for each substance are displayed.
According to Eq.(\ref{eq:v_E}), the differences in mobility allows to
correlate time with position taken by an ion which arrives on a detector.  In
order to explain experimental results, we need to consider that the mobility
of Toluene ($K_0 = 1,87$) is lower than that of Ethanol ($K_0 = 2,06$) and
Acetone ($K_0 = 2,13$)~\cite{Jurs(1996),Agbonkonkon(2004)} and that with the
configuration proposed, the electric field is perpendicular to the direction
of the carrier gas (see Fig.~\ref{fig:barrido}).  Hence, Toluene ions should
travel longer distances without being deviated compared with Ethanol and
Acetone ions. For these last ones, the scarce relative difference of 3.5\%
reported between the Acetone (larger) and Ethanol (smaller)
mobilities~\cite{Jurs(1996), Agbonkonkon(2004)}, might yields to similar path
deflection and nearly the same distance of impact on detectors produced by the
swept-field scanning. In general, it is expected that lighter ions have
larger mobilities as they have smaller inertial mass. Although this rule
applies for a large number of molecules~\cite{Jurs(1996),Agbonkonkon(2004)}
there are some exceptions where the opposite is true. Such exceptions can be
explained in terms of a fragmentation process of the primary ion to lighter
secondary ions and, therefore, with greater mobility. Due to the type of
ionization proposed in this work, we suppose that the fragmentation of Acetone
and Ethanol produce secondary ions with highly relative
abundance~\cite{NistAcetona(2007)}.  Consequently, each species aquire a
different mobility compared to that expected in terms of primary ions mass
alone.

\subsection{Times of Flight}\label{subsec:tof}

The velocity of the carrier gas and the difference of electric potential
within the cell can be conveniently combined in order to modulate the amount
of ions that goes to a given place of the experimental device. We shall
analyse here the time of flight for each ion and the average rate of the
collected charge in each detector as a function of both the velocity of the
carrier gas and the difference of electric potential.
\begin{figure}[h]
\begin{center}
\includegraphics[height=6cm, bb =50 50 301 226]{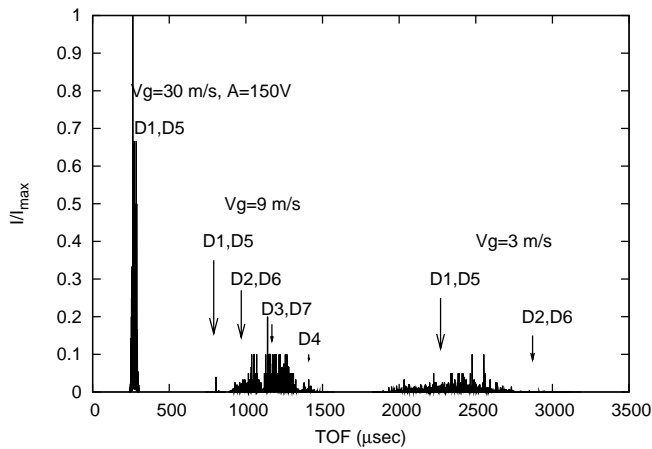}
\caption{\label{fig:tof2} Normalized $I/I_{max}$ current intensity vs. time of
  flight (TOF) for each detector $D_i$ with $i=1\ldots7$ ordered by pairs as
  shown in the graph labels. For $v_g =$ 3 and 9 m/s, the difference of
  potential is $A =$ 15 V.%
}
\end{center}
\end{figure}
In Fig.~\ref{fig:tof2} we display the normalized current generated in each
detector by the ions of Toluene when the velocity of the carrier gas is 3, 9
and 30 m/s and the differences of electric potential in each detector cell are
15 and 150 V. A pattern that spreads the current over time of flights when
both the velocities $v_g$ and potentials diminishes is observed. This can be
explained considering that the contribution to the ion's energy from the
interaction between the ions and the carrier gas do compete with the
electrical counterpart. Increasing the electric potential differences of the
cell will give a larger kinetic energy contributtion to the ions compared with
the contribution from the Brownian motion. Thus, such electric potential
increment clearly produce a lesser diffuse and so more defined and streched
ion paths. For that reason, in Fig.~\ref{fig:tof2} it is observed that, as
both the velocity of the carrier gas and the potential decrease, the average
rate of ions that arrive at the same time also decrease. This means that in
such case the ions arrival are expanded over the whole detection zone.

\subsection{Collected charges in detectors}\label{subsec:ca}

Counting the amount of collected charge in each detector offers an alternative
way to get the times of flight. Their results are also in agreement with
experimental data. We present here the measurement of the collected charge
when the velocity of the carrier gas is 30 m/s and the difference of potential
in the cell is 50 V.
\begin{figure}[h]
\begin{center}
\includegraphics[height=6cm, bb = 50 50 266 201]{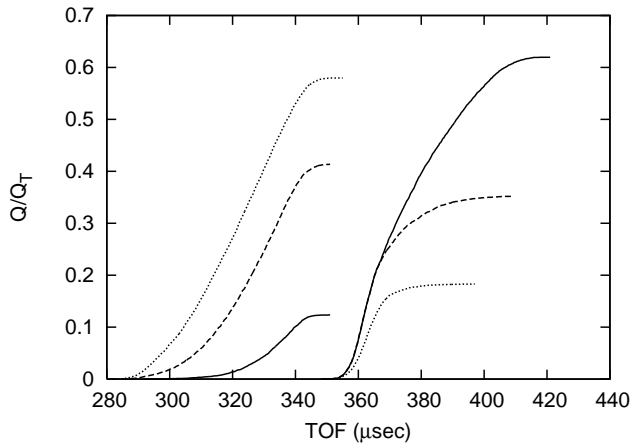}
\caption{\label{fig:TolAcetEthaQvsT} Normalized total charge $Q/Q_T$ for
  detectors $D_6$ (left) and $D_7$ (right) for each species as a function of
  time of flight when primary ions were considered for Toluene (full lines),
  Ethanol (dotted lines) and Acetone (dashed lines).%
}
\end{center}
\end{figure}
It is verified that the times of flight are larger for ions that arrive on
detectors which are farther to cell inlet. For instance, in
Fig.~\ref{fig:TolAcetEthaQvsT} the same graph for detectors $D_6$ and $D_7$ is
shown for the collected charge $Q$ divided by total charge $Q_T$ of ions when
Toluene, Acetone and Ethanol are tested separately. 

When mobility is estimated from ion mass, as should be if we consider the case
of fragmentation process during the ions generation, the charge collected in
each detector as a function of time of flight follows the simple rule expected
by inertial mass considerations. This can be observed in Fig.
~\ref{fig:TolAcetEthaQvsT}. The collected charge for Ethanol is greater than
that of Acetone for times of flight between 280 and 350$\mu$s corresponding to
the arrival of ions at the detector $D_6$.  However, if we consider the
mobilities reported in the literature we could expect that this result should
be the opposite one. For that reason, we performed our simulation also
considering groups of secondary ions. We supposed that when Acetone, Ethanol
and Toluene were ionized they would form groups of ions from the fragmentation
of the primary ion.  We considered the ions with a relative abundance greater
than 10\%. Furthermore, using secondary ions, we found out a large defocusing
and a low temporal definition due to a wide charge distribution on the
detectors assembly. Then as an improvement we considered more appropriate to
represent the collected charges in each detector taking care also of the
secondary ions for each species.
\begin{figure}[h]
  \begin{center}
    \includegraphics[height=6cm]{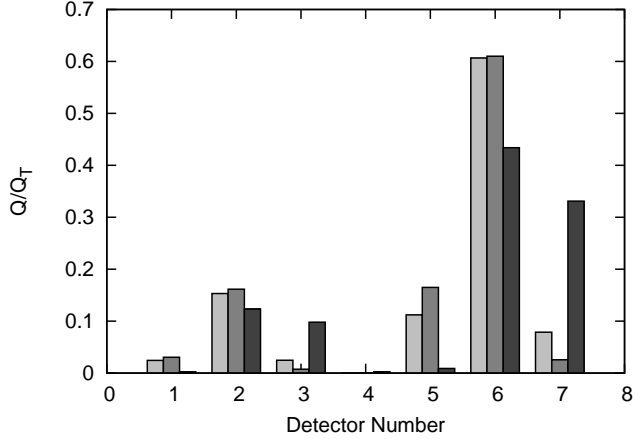}
    \caption{\label{fig:qvsdet} Normalized collected charge $Q/Q_T$ in each
      detector obtained from the fragmentation product of the primary ion with
      relative abundance greater than 10\% for Acetone (light gray), Ethanol
      (gray) and Toluene (dark gray). Over 6000 ion paths are averaged for
      each species considering carrier gas velocities between 20 and 30
      m/s. Numeration and configuration of detectors is the same as that of
      the experimental device.%
    }
\end{center}
\end{figure}
In Fig.~\ref{fig:qvsdet} we display such results. In this case a charge
distribution closer to that expected in terms of Acetone ion mobilities is
obtained. However, a larger amount of Ethanol ions than that of Acetone ions
in detectors ($D_1$,$D_5$) is still observed. Such detectors are the first
pair from the entrance to the cell of the carrier gas. The second pair of
detectors is ($D_2$,$D_6$). For this second pair, the method of secondary ions
seems to be more consistent with the expected result in terms of known
mobilities. The detector pair that follows is ($D_3$,$D_7$). In this last
case, the result that arise when considering fragmentation is again opposite
to that which would be expected from mobilities.

\subsection{Principal components analysis}\label{subsec:pca}

Based on the previous background on simulation and experimental results we are
be able do another step forward. Our working assumption is now that once the
cell geometry and the operational conditions of our device are given (the
carrier gas flux and the cell potential values) the charge collected in each
detector has its own statistics for each unknown species. Specifically, the
analysis of the principal components applied to our device allows us to
determine the fingerprint of each species corresponding to our experimental
configuration.  We associate a random variable $X^i$ ($i=1\ldots7$) to the
collected charge $Q$ in each detector, normalized with the total charge $Q_T$
and distributed on all the detectors. Thus, $X^i$ are real numbers in the interval
[0, 1]. Each measurement of the collected charge in a detector corresponds to
an event of the random variable $X^i$. Therefore, if we perform $N$
measurements for each variable, we can construct the vector $\vec
X^i=(X^i_1,X^i_2,\ldots,X^i_N)^T$ whose coordinates are the result of these
measurements. The notation $(\ldots)^T$ means the transpose of
$(\ldots)$. From the definition of the mean value $\bar X^i=1/N\sum_l X^i_l$ and
experimental covariance~\cite{Meyer(1986)}, a real symmetric and covariant
matrix ${\mathbf C}$ is obtained, whose dimension is equal to the number of
detectors, {\it i.e.} 7$\times$7. Their elements are defined by
\begin{equation}
  \label{eq:cov}
  {\mathbf C}_{ij}\equiv Cov(X^i,X^j)=\frac{1}{(N-1)}\sum_m(X^i_m-\bar X^i)(X^j_m-\bar X^j),
\end{equation}
where $m=1\ldots N$. We now introduce the similarity transformation
\begin{equation}
  \label{eq:P}
  \mathbf D=\mathbf P^T \mathbf C \mathbf P,
\end{equation}
where $\mathbf D$ is a diagonal matrix and $ \mathbf P$ is an orthogonal
matrix containing the eigenvectors of $\mathbf C$ ordered by columns. From the
last matrix it is possible to define the transformation
\begin{equation}
  \label{eq:Y7}
  \mathbf Y={\mathbf P^T} \mathbf X, 
\end{equation}
That enable us to change from random variables of detectors
\begin{equation}
  \label{eq:X}
  \mathbf X = (\vec X^1,\vec X^2,\ldots,\vec X^7)^T.
\end{equation}
to a new set of random variables
\begin{equation}
  \label{eq:Y}
  \mathbf Y = (\vec Y^1,\vec Y^2,\ldots,\vec Y^7)^T,
\end{equation}
From this output a principal subgroup can be obtained 
\begin{equation}
  \label{eq:Z}
  \mathbf Z={\mathbf Q^T} \mathbf X, 
\end{equation}
being ${\mathbf Q}$ the matrix formed by the first $n\le7$ eigenvectors of $\mathbf C$ ordered in decreasing order of its eigenvalues. The final results

\begin{equation}
  \label{eq:Zn}
  \mathbf Z = (\vec Z^1,\ldots,\vec Z^n)^T,
\end{equation} 
contain the expected principal components. Different criteria were proposed
for getting $n$ ~\cite{Pearson(1901)}, although all of them consider the minor
subspace in which the greatest quantity of data is
concentrated~\cite{Lamagna(2004)}.

The results of Eq.(\ref{eq:Y}) can be understood in terms of a coordinate
transformation. Within the new coordinate system, each component is
independent. This allows us to define an orthogonal space that represent the
data obtained from measurements of collected charges for each species. The
principal eigenvectors used show in turn the specific combination of the
original detectors that intervene in the output of the measurement.
\begin{figure}[h]
  \centering
  \includegraphics[height=6cm]{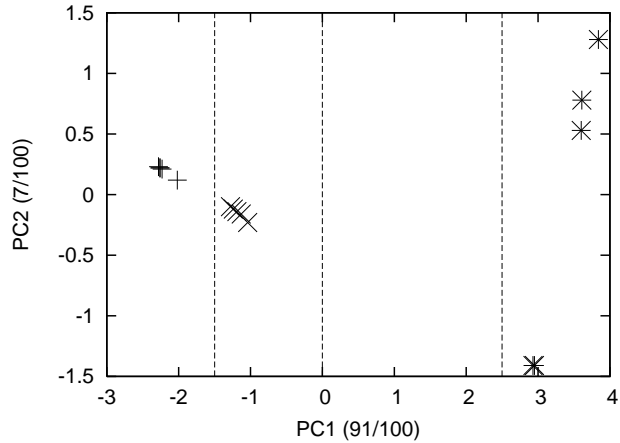}
  \caption{\label{fig:pca7det} The first two principal components for 5
    measurements of the charge collected in each detector when using all of
    them. Ethanol (plus symbol), Acetone (cross symbol) and Toluene (asterisk
    symbol) averaged on 6000 runs each species.%
  }
\end{figure} 

\begin{figure}[h]
  \centering
  \includegraphics[height=6cm]{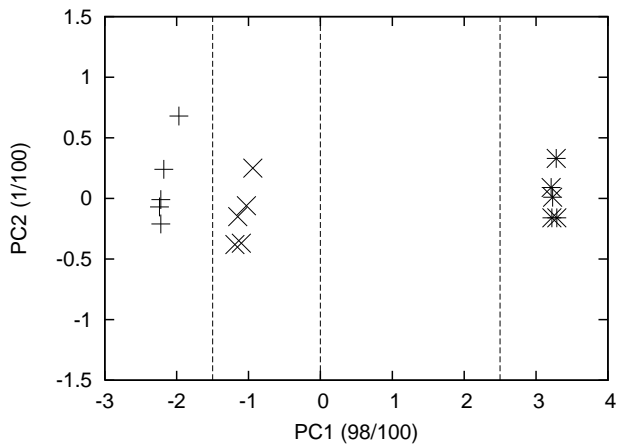}
  \caption{\label{fig:pca6det} The first 2 principal components for 5
    measurements of collected charge in each detector, but without detector
    number 4. Ethanol (plus symbol), Acetone (cross symbol) and Toluene
    (asterisk symbol) averaged on 6000 runs each species.
  }
\end{figure}

Fig.~\ref{fig:pca7det} shows the coordinates of the two first principal
components for a set of 5 measurements of the collected charge in each
detector for Ethanol, Acetone and Toluene ions and fragments of them. Each one
of these measurements is similar to that represented in Fig.~\ref{fig:qvsdet}
and corresponds to the average values on a sample of 6000 ion paths,
considering the carrier gas velocities between 20 and 30 m/s. Only 2\% of data
remains outside the analyzed set by the two first principal components,
distributing 91\% in the first and 7\% in the second of
them. Fig.~\ref{fig:pca6det} shows the coordinates
in the two first principal components of measurements on the same system
considered in Fig.~\ref{fig:pca7det}, but with one variable less due to
detector number 4 was not considered. In this case it is observed that 99\% of
data is represented by the first two components and 98\% only by the first
one. This is explained by the fact that being very scarce the collected
charges on detector 4, as can be seen in Fig.~\ref{fig:qvsdet}, the diminution
of the number of variables decreases the dimension of the total space
considered with a very little alteration of the sample size. Therefore, the
concentration of data in the first two principal components increase.  The
relevance of the PCA result is considerable regarding the separation of the
three set of data for all species we have studied. The identification of an
unknown sample is positive if its representation fits within any of the
clearly non-overlapping region of our three set of data (see
Figs.~\ref{fig:pca7det} and \ref{fig:pca6det}).

\section{Conclusions}\label{sec:conclusion}

Our simulation clearly shows that for the trace of ions, the carrier gas
velocity and the cell potential are directly correlated. The analysis of
fragmented ions partially explain the traces of Acetone and Ethanol when their
mobilities are estimated from fragments mass. However, the integrated signals
which were obtained experimentally show the same trend that our simulation. As
shown in Fig.~\ref{fig:DetAllExp}, when comparing the signals of detectors
($D_1$,$D_5$) with the subsequent pair ($D_2$,$D_6$), it is observed that they
maintain almost the same proportion between all substances for each detector,
but the whole amount is increased in the second pair. In the third pair,
($D_3$,$D_7$), the proportion of the trend between substances almost remains
the same, but decreasing the total amount of the set. Only in the most distant
detector the result is different and shows the proportions between species
following the expected results in terms of tabulated
mobilities~\cite{Jurs(1996),Agbonkonkon(2004)}.

Since the simulations of the times of flight for each substance shows a
characteristic distribution for both, charges and currents obtained through
the detectors assembly, i.e. a {\em fingerprint}, the statistical correlation
in the distribution of collected charges in the detectors using the Principal
Component Analysis can be applied.

For the experimental prototype proposed here a set of specific parameters
(velocity of carrier gas and electric potential of the cell) to define
Acetone, Ethanol and Toluene traces were optimized by numeric simulation.
Afterwards a PCA statistical analysis of the collected ion charges in the set
of detectors is used to define a non-overlapping region in a appropriated
diagram for each of the above mentioned compounds.  Thus, our procedure, which
involves a training to develop the proper fringerprint of each compound, can
be used in the designing of devices for the identification with certainty of
unknown compounds,

\section{acknowledgment}
GPO acknowledge illuminating discussions with W. Luis Moch\'an and useful
comments of Patricio F. Provasi and Rodolfo H. Romero. This work was supported
by FONCyT (grant PAE 22592/2004 nodo NEA:23016 and nodo CAC:23831).


\begin{thebibliography}{10}

\bibitem{Moore(2004)}
D.S. Moore.
\newblock Instrumentation for trace detection of high explosives.
\newblock {\em Rev. Sci. Instrum.}, 75:2499, 2004.

\bibitem{Revercomb(1975)}
H.E. Revercomb and E.A. Mason.
\newblock Theory of plasma chromatography/gaseous electrophoresis- a review.
\newblock {\em Anal. Chem.}, 47:970, 1975.

\bibitem{Jurs(1996)}
M.D. Wessel, J.M. Sutter, and P.C. Jurs.
\newblock Prediction of reduced ion mobility constants of organic compounds
  from molecular structure.
\newblock {\em Anal. Chem.}, 68:4237, 1996.

\bibitem{Agbonkonkon(2004)}
N.~Agbonkonkon, H.D. Tolley, M.C. Asplund, E.D. Lee, and M.L. Lee.
\newblock Prediction of gas-phase reduced ion mobility constants ($k_0$).
\newblock {\em Anal. Chem.}, 76:5223, 2004.

\bibitem{Appelhans(2005)}
A.D. Appelhans and D.A. Dahl.
\newblock Simion ion optics simulation at atmospheric pressure.
\newblock {\em Int. J. Mass Spectrom.}, 244:1, 2005.

\bibitem{Eiceman(1999)}
J.I. Baumbach and G.A. Eiceman.
\newblock Ion mobility spectrometry: Arriving on site and moving beyond a low
  profile.
\newblock {\em Appl. Spectrosc.}, 53:338A, 1999.

\bibitem{Miller(2000)}
R.A. Miller, G.A. Eiceman, E.G. Nazarov, and A.T. King.
\newblock A novel micromachined high-fiel asymmetric waveform-ion mobility
  spectrometer.
\newblock {\em Sens. Act. B Chem.}, 67:300, 2000.

\bibitem{Gillig(2004)}
K.J. Gillig, B.T. Ruotolo, E.G. Stone, and D.H. Russell.
\newblock An electrostatic focousing ion guide for ion mobility-mass
  spectrometry.
\newblock {\em Int. J. Mass Spectrom.}, 239:43, 2004.

\bibitem{Solis(2006)}
A.A. Solis and E.~Sacrist\'an.
\newblock Designing the measurement cell of a swept-field differential
  aspiration condenser.
\newblock {\em Rev. Mex. Fis.}, 52:322, 2006.

\bibitem{Zimmermann(2007)}
S.~Zimmermann, N.~Abel, W.~Baether, and S.~Barth.
\newblock An ion-focusing aspiration condenser as an ion mobility spectrometer.
\newblock {\em Sens. Actuators B}, 125:428, 2007.

\bibitem{Lai(2008)}
H.~Lai, T.R. Mcjunkin, C.J. Miller, J.R. Scott, and J.R. Almirall.
\newblock The predictive power of simion/sds simulation software for modeling
  ion mobility spectrometry instruments.
\newblock {\em Int. J. Mass Spectrom.}, 2008.
\newblock to be published.

\bibitem{Pearson(1901)}
K.~Pearson.
\newblock On lines and planes of closest fit to systems of points in space.
\newblock {\em Philosophical Magazine}, 2:559, 1901.

\bibitem{Lamagna(2004)}
A.~Lamagna, S.~Reich, D.~Rodriguez, and N.N. Scoccola.
\newblock Performance of an e-nose in hops classification.
\newblock {\em Sens. Actuators B}, 102:278, 2004.

\bibitem{ErrataSDS}
The points of the cube considered by the code SIMION 8.0/SDS in the line 1044
  of {\em collision\_sds.lua} define directions to their corners more likely.
  Thus, for isotropic case such line code is not correct for local definition
  of random direction to be model in a step movement on the path of ions.

\bibitem{Nist(2005)}
P.J. Linstrom and W.G. Mallard, editors.
\newblock {\em NIST Chemistry WebBook}.
\newblock Number~69 in Nist Standard Reference Database. National Institute of
  Standard and Technology, Gaithersburg MD, 20899, 2005.

\bibitem{NistAcetona(2007)}
Japan AIST/NIMC Database-Spectrum MS-IW-7880.
\newblock {\em Acetone}.
\newblock NIST Mass Spectrometry Data Center, USA, 2007.

\bibitem{Meyer(1986)}
P.Meyer.
\newblock {\em Probabilidad y aplicaciones estadisticas}.
\newblock Addison-Wesley Iberoamericana, 1986.

\end{thebibliography}

\end{document}